\documentstyle[twocolumn,aps]{revtex}
\begin{document}
\draft
\twocolumn[\hsize\textwidth\columnwidth\hsize\csname @twocolumnfalse\endcsname
%
%

\title{ Electronic Hamiltonian for Transition Metal Oxide Compounds}


\author{ Erwin M\"uller-Hartmann$^1$  and Elbio Dagotto$^2$}

\address{$1.$    Institut f\"ur Theoretische Physik, Universit\"at zu K\"oln, 
   Z\"ulpicher Str. 77, D-50937 K\"oln, Germany.  }
\address{$2.$ Department of Physics and National High Magnetic Field Lab,
Florida State University, Tallahassee, FL 32306, USA.}

\date{\today}
\maketitle

\begin{abstract} 
An effective electronic Hamiltonian for transition metal oxide
compounds is presented. For Mn-oxides, the Hamiltonian
contains spin-2 ``spins'' and spin-3/2 ``holes'' as degrees of freedom.
The model is constructed from the Kondo-lattice Hamiltonian for mobile $e_g$ 
electrons and 
localized $t_{2g}$ spins, in the limit of a large Hund's coupling. 
The effective electron bond hopping amplitude
fluctuates in sign as the total spin of the bond changes. In the large
spin limit, the hopping amplitude for electrons aligned with the core ions
is complex and a Berry phase is accumulated when these electrons move in 
loops.
The new model is compared with the standard double exchange Hamiltonian.
Both have ferromagnetic ground states at finite hole density and low
temperatures, but 
their critical temperatures could be substantially different due to
the frustration effects induced by the Berry phase.

\end{abstract}

\vskip2pc]
\narrowtext

%
%

The discovery of giant magnetoresistance effects in ferromagnetic
metallic oxides ${\rm R_{1-x} X_x Mn O_3}$ $({\rm where~ R=}$
${\rm La,Pr,Nd;~X=Sr,Ca,Ba,Pb})$ has
triggered renewed attention into these compounds.\cite{jin}
A decrease in resistivity of four orders of magnitude
has been observed in thin films of ${\rm Nd_{0.7} Sr_{0.3} Mn
O_3}$ at fields of $\sim 8T$.\cite{xiong} The phase diagram of ${\rm
La_{1-x} Ca_x Mn O_3}$ is very rich with ferromagnetic (metal and insulator) phases, as
well as regions where charge-ordering 
is observed.\cite{phase} The magnetic and electronic
properties of these manganese oxides are believed to 
arise, at least in part, 
from the strong coupling between correlated itinerant electrons and localized
spins, both of $3d$ character. The ${\rm Mn^{3+}}$ ions 
have three electrons in the $t_{2g}$ state forming a local
S=3/2 spin, and one electron in the $e_g$ state which hops between
nearest-neighbor Mn-ions, with double occupancy suppressed by Coulombic
repulsion.
The widely used Hamiltonian to describe manganese oxides is\cite{anderson}
$$
{H = -t \sum_{\langle {\bf m}{\bf n} \rangle \sigma} (c^\dagger_{{\bf m}
\sigma} c^{\phantom{\dagger}}_{{\bf n} \sigma} + H.c.) - J_{H} 
\sum_{\bf n} {{\bf\sigma}_{\bf n}}\cdot{{\bf S}_{\bf n}}, }
\eqno(1)
$$
\noindent where the first term is the $e_g$
electron transfer between nearest-neighbor Mn-ions at sites ${\bf m,n}$, while the second
term is the ferromagnetic Hund coupling between the S=3/2 localized spin
${\bf S_{n}}$ and
the mobile electron with spin ${\bf \sigma_{n}}$ ($J_{H} > 0$). 
A Coulombic repulsion to suppress double occupancy in the itinerant band is implicit.
The on-site Hund coupling energy is
larger than the conduction bandwidth favoring the alignment
of the itinerant and localized spins. For ${\rm Mn^{3+}}$, the resulting 
spin is 2, while for ${\rm Mn^{4+}}$ (vacant $e_g$ state) the spin is 3/2.

Since the study of Hamiltonian Eq.(1) is a formidable task,
simplifications have been introduced to analyze its properties. 
A familiar approach is the use of the double-exchange 
Hamiltonian,\cite{anderson,zener,gennes,furukawa,inoue,millis} 
where the $e_g$ electrons
move in the background of classical 
spins ${\bf S}^{cl}_{\bf n}$ that approximate the
S=3/2 almost localized $t_{2g}$ electrons. The conduction electron
effective hopping between sites ${\bf m}$ and ${\bf n}$
used in previous work is
$t^{eff}_{{\bf m}{\bf n}} = t \sqrt{ 1 + 
({{\bf S}^{cl}_{\bf m} }\cdot{{\bf S}^{cl}_{\bf n}}/S^2) } $, 
where $S$ is the magnitude of the classical spin.\cite{anderson,gennes}
Using this model, the ferromagnetic critical temperature, $T_c$, was recently
estimated.\cite{millis} Since the result was much larger than experimentally 
observed, the need for non-electronic interactions to describe 
Mn-oxides was remarked.\cite{millis,hwang}

The purpose of this
paper is to reexamine the large $J_{H}$ limit of model Eq.(1).
We derive an effective Hamiltonian for Mn-oxides which is valid for the
quantum mechanical case of S=2 ${\rm Mn^{3+}}$ ions. Actually, the
Hamiltonian is discussed 
for an arbitrary spin S. The ideas followed in this paper
are a generalization of the calculation recently presented for NiO
compounds, where ``holes'' doped into a S=1 background carry S=1/2
and they move following nontrivial hopping processes.\cite{nio}
At large S, the model described here contains a
complex effective coupling for electrons whose spin is aligned with the
core spins. These electrons acquire a phase when they move in
closed loops, an effect not taken into account in previous
literature for Mn-oxides.  Although the double-exchange model is 
not recovered at large S,
the presence of a ferromagnetic phase at low temperatures
is likely even in the revised model and most of the differences between
double-exchange and the new model will occur at finite temperatures.

In the limit $J_{H} \rightarrow \infty$, it is natural to restrict
 the Hilbert space corresponding to a given site   to
spin eigenstates with the maximum allowed spin $S'$, 
and projection $m'=-S',...,S'$, compatible
with the number of electrons at that site.
Thus, the ions are in a state
either with $S'=S$, if the $e_g$ level is occupied, or
of $S'=S-1/2$, if there is no $e_g$ electron. The corresponding 
degrees of freedom will be referred to as ``spins'' and 
``holes'', respectively. They are coupled by standard nearest-neighbor
Heisenberg interactions, plus ``hopping'' terms 
for the movement of holes, respecting the large
$J_{H}$ approximation.
The site states will be denoted by $|S',{m'} \rangle_{\bf n}$. 
For charge-transfer compounds, the absence of an $e_g$ electron
can be considered as 
caused by the Zhang-Rice singlet formation between oxygen and $d$
electrons.\cite{nio} 
Thus, our results are not restricted to Mott-Hubbard compounds but are valid for
transition metals in general in arbitrary dimensions, and contain as a special case
the well-known t-J model widely used for cuprates.

The hopping amplitude $t_{{\bf m}{\bf n}}$ for nearest-neighbor sites
${\bf m}$ and ${\bf n}$
certainly depends on the values of the spin at both sites, as well as on their
projections.  
Let us assume that at site ${\bf m}$ we have a spin $S'=S$, while 
at site ${\bf n}$ there is a $S'=S-1/2$ hole. 
The eigenstates of the ${\bf mn}$ bond can be labelled by the total bond
spin, $S_T$, and its projection, $M_T$, and they admit an expansion in the
basis compatible with $J_{H} \rightarrow \infty$ as
$$
{ {\widetilde {| S_T, M_T \rangle } }_{i} 
= \quad\quad\quad\quad\quad\quad\quad\quad\quad\quad\quad\quad\quad\quad
}
$$
$$
{ =\sum_{m m'} 
| S, m \rangle_{\bf m}
| S-{{1}\over{2}}, m' \rangle_{\bf n} C(S,S-{{1}\over{2}},m,m',S_T,
M_T ),
}
\eqno(2)
$$
\noindent where $C$ are Clebsch-Gordan (CG) coefficients, and
 $m$ ($m'$) runs
from $-S$ to $+S$ ($-(S-{{1}\over{2}})$ to $+(S-{{1}\over{2}})$). Let us consider Eq.(2) as the
$initial$ state, i.e. the one before the hopping occurs.
The $final$ state $ {\widetilde{ | S_T, M_T
\rangle_{f}} }$, where the hole has moved from ${\bf n}$ to ${\bf m}$,
admits the same decomposition but with the site indices permuted. Since
the hopping Hamiltonian $H_t$ (first term in Eq.(1)) 
is a scalar, the matrix element necessary
to evaluate the effective hopping amplitude is
$$
{ t(S_T,M_T) = \widetilde{ _{f}{\langle S_T M_T |}} H_t {\widetilde {
| S_T M_T \rangle_{i} }} .
}
\eqno(3)
$$
\noindent Using the Wigner-Eckart theorem and after long, but
straightforward, CG algebra it can be shown that
$$
{ t(S_T,M_T) = t {{S_T + 1/2}\over{2S}} (-1)^{2S - S_T - 1/2}.
}
\eqno(4)
$$
\noindent The amplitude of this hopping, i.e. $|t(S_T,M_T)|$, 
is in agreement with the well-known results used by Zener,\cite{zener}
Anderson and Hasegawa,\cite{anderson} and
de Gennes\cite{gennes} to introduce the double-exchange model. 
Actually, it can be easily shown that for any value of the spin 
the following operatorial equality holds:
$$
{
{{S_T + 1/2}\over{2S}} = {{1}\over{\sqrt{2}}} 
\sqrt{ 1 + {{1}\over{2S}} + {{{\bf S}_{\bf m}}\cdot{{\bf S}_{\bf n}}\over{S^2}}      },
}
\eqno(5)
$$
\noindent where ${\bf S}_{\bf m},{\bf S}_{\bf n}$ are spin operators 
that can act on a spin or a hole.
However, note the presence of a nontrivial
$S_T$-dependent sign in the effective hopping Eq.(4).\cite{sign}
We will argue here that the presence of this sign is crucial for a
proper quantum mechanical treatment of the large $J_{H}$ limit of Eq.(1).

To better understand this nontrivial sign,
consider, e.g., the special case of Eq.(1) where the localized 
degree of freedom has
spin 1/2. Thus, at the link ${\bf mn}$ used in Eqs.(2-4), and working
at large $J_{H}$, the problem reduces to two electrons forming a
 spin $S'=1$ in one site,
interacting with one 
electron (i.e. a $S'=1/2$ hole) at the other site. 
Let us verify Eq.(4) for some 
special cases using Eq.(1): (1) If $S_T =3/2$ and
$M_T=3/2$, moving the $e_g$ electron across the ${\bf mn}$ bond
produces a matrix element 1, in agreement with
Eq.(4); (2) Suppose now that $S_T =1/2$ and $M_T=1/2$.
Using CG coefficients, the initial state before the hopping occurs
can be written as
$ {\widetilde{ |1/2,1/2\rangle_{i} }}
=({1}/{\sqrt{3}}) | 1/2,1/2\rangle_{\bf m}
|1,0\rangle_{\bf n}
-(\sqrt{{2}/{3}}) | 1/2,-1/2\rangle_{\bf m} |1,1\rangle_{\bf n},
$
\noindent in the basis where at site ${\bf m}$ there is a $S=1/2$ hole and
at ${\bf n}$ a $S=1$ spin. Applying explicitly the hopping Hamiltonian $H_t$ we
obtain
$ H_t {\widetilde{ |1/2,1/2 \rangle_{i}} } =$ $ -({t}/{2})
{ {\widetilde{ |1/2,1/2 \rangle_{f} } } }
- ({t \sqrt{3}}/{2}) |0,0\rangle_{\bf m} |
1/2,1/2 \rangle_{\bf n}$.
The last term is not  favored by the strong Hund's coupling,
and thus the  matrix element relevant for the low energy effective
Hamiltonian is simply ${\widetilde {_{f}\langle 1/2, 1/2|}} 
H_t {\widetilde { | 1/2,1/2 \rangle_{i}}} = - t/2$.
The amplitude has the expected absolute value $|t(S_T,M_T)|$
compatible with the double exchange model.
However, the matrix element is actually $negative$,
compatible with our result Eq.(4). The negative sign originates in
${\widetilde { | 1/2,1/2 \rangle_{i}}},$ where in order to make a total spin 1/2 combination out of
individual spins 1/2 and 1, amplitudes of different signs are needed.
While such effects
are natural in quantum mechanical processes involving state overlaps,
they  are not included in the double-exchange model.
These signs are important to properly reproduce
the physics of model Eq.(1) at large $J_{H}$ even close to a
ferromagnetic configuration since just small deviations from a fully
polarized link can involve a change in the sign of the hopping. 
If the sign is included, the hopping amplitude mean value
vanishes at large spin $S$, while it is finite in the double exchange
approximation.

While the $S_T$ basis Eq.(2) is useful to find the hopping amplitudes,
the effective Hamiltonian for the complete lattice has a more intuitive form
in the basis of spin eigenstates at each site. Using Eq.(5)
we can write an  effective hopping Hamiltonian as a polynomial
$$
{  H_{eff} = -t \sum_{ \langle {\bf m}{\bf n} \rangle} P_{{\bf m}{\bf n}}
~Q_{S}(y),
}
\eqno(6)
$$
\noindent where ${ P_{\bf mn} }$ is an operator that permutes the states
at sites
${\bf m}$ and ${\bf n}$, 
$y = {{{\bf S}_{\bf m}}\cdot{{\bf S}_{\bf
n}}}/ S (S-1/2)$, and ${\bf S_m,S_n}$ acts 
either on the localized spin or the hole 
(actual spins $S$ and $S-1/2$, respectively), as explained before.
$Q_S(y)$ is spin-$S$ dependent and it can be found iteratively
starting from the lowest spin case using the relation
$$
Q_S(y) = - a_S Q_{S-1/2}
[ b_S (2+(2S-1)y)/(2S+1) ] +
$$
$$
{+ [1+ a_S Q_{S-1/2} ( { b_S}   )] 
\prod_{l=1}^{2S-1} { {S(2S+1+(2S-1)y)-l^2 }\over{ 4S^2-l^2}},}
\eqno(7)
$$
where we used $Q_{1/2}(y) = 1$, $a_S = (2S-1)/2S$, and
$ b_S = S(2S+1)/[(S-1)(2S-1)]$. 
For $S=1$, this implies 
 $Q_1(y) = (1+y)/2$ which
correctly reproduces the Hamiltonian recently derived for doped
${\rm Y_2 Ba Ni O_5}$ compounds.\cite{nio} 
If only the absolute value of the matrix
elements Eq.(3) would have been used, then $Q_1(y)$ becomes instead
$(5+y)/6$. This appa\-rent\-ly small difference is
nevertheless of
much relevance: in the proper $Q_{S=1}(y)$ polynomial,
the hopping 
$|1/2,-1/2\rangle_{\bf m} |1,1\rangle_{\bf n}$ 
$\rightarrow$ $|1,1\rangle_{\bf m}
|1/2,-1/2\rangle_{\bf n}$ has zero amplitude,
compatible with Hamiltonian Eq.(1) where the spin projection at each
site can only change in units of 1/2,
after one $e_g$ electron moves. 
However, if the hopping amplitudes signs
are neglected, such unphysical processes become 
incorrectly allowed. Such hidden sum-rules
of the Kondo-lattice model are also illustrated when the
effective Hamiltonian acts over an arbitrary link state. The result is
$ H_{eff}|S,m \rangle_{\bf m} |S-1/2,m' \rangle_{\bf n} = 
-t [ A(S,m ,m') |S-1/2,m+1/2\rangle_{\bf m}
|S,m'-1/2\rangle_{\bf n} + 
B(S,m,m')
|S-1/2,m-1/2\rangle_{\bf m} |S,m'+1/2\rangle_{\bf n} ],
$
where $A(S,m,m') = {\sqrt{(S-m'+1/2)(S-m)}/ 2S} $
and $B(S,m,m') = {\sqrt{(S+m'+1/2)(S+m)} / 2S}$,
clarifying the hopping processes that are allowed. 

For the particular case of S=2, which would be applicable to
${\rm R_{1-x} X_x Mn O_3}$, the polynomial becomes
$$
{ Q_{S=2}(y) = -1 - {{5}\over{4}} y + {{7}\over{4}} y^2 + {{3}\over{2}} y^3 .
}
\eqno(8)
$$
Thus, we propose  Eqs.(6,8) 
as the effective Hamiltonian for Mn-oxides 
in the large ${ J_{H}}$ limit.\cite{spin} It is the 
generalization to Mn-oxides of the t-J model for Cu-oxides.
Ours is a fully quantum mechanical
model for spin 2 ${\rm Mn^{3+}}$ that takes into account the proper signs of the
effective hopping amplitudes. Unlike the
double exchange model, there are no classical spins in our effective
model. However, note that for a small exchange $J$ 
among the spins (in our approach there is always an implicit 
Heisenberg coupling between ions) and a small hole density,
model Eq.(8) favors a ferromagnetic ground state to improve the hole
kinetic energy, as the double exchange model does. Recent 
many-body numerical results
applied to the case of S=1 in one dimension\cite{nio} have indeed
detected a robust ferromagnetic state at zero temperature for a wide
range of densities.\cite{jose}

To gain further intuition about the relevance of the signs in the
effective hopping Eq.(4), we studied the Kondo-like
Hamiltonian Eq.(1) for the special case where at large Hund coupling the 
localized spins are classical. Note that 
it is not well justified to apply this limit to Manganites,
since the neglected spin flip processes have 
amplitude $1/\sqrt{2S}$ (see Eq.(5)) which is not small for $S=2$.
Nevertheless, to establish a connection with the double exchange model
it is necessary to work at large $S$.
In this limit, it is
convenient to rotate the $e_g$ electrons such that their spin
quantization axis is parallel to the core spins.\cite{millis} 
Since the Hund's coupling is large, in the rotated basis
only the ``spin-up'' component of the hopping matters, 
and the nontrivial effects of the core spins appear in
the modulation of the hopping. In this
limit the Hamiltonian becomes
$$
{H = -\sum_{\bf \langle lm \rangle }(t^{\phantom{\dagger}}_{\bf lm} 
d^{\dagger}_{\bf l} d^{\phantom{\dagger}}_{\bf m}+h.c)},
\eqno(9) 
$$
where $d_{\bf n} = \cos(\theta_{\bf n}/2)c_{\bf n\uparrow}+
i\sin(\theta_{\bf n}/2)\exp(i\phi_{\bf n})c_{\bf n\downarrow} $ 
are rotated electron operators with spin up, and the hopping amplitude is a 
complex number given by
$$
t_{\bf  lm } = \cos({{\theta_{\bf l}}\over{2}})\cos({{\theta_{\bf
m}}\over{2}}) +\sin({{\theta_{\bf l}}\over{2}})\sin({{\theta_{\bf m}}\over{2}})
\exp[i(\phi_{\bf m}  - \phi_{\bf l} )].
\eqno(10)
$$
When electrons move with this nontrivial hopping they effectively
collect a 
phase. Alternatively, the problem can be rephrased as that of electrons
moving with hopping $|t_{\bf lm}|$ in the presence of a nonuniform
gauge field $e^{i{\bf A_{lm}}}$
which is the phase of the hopping.\cite{doucot}
This phase can have important interference consequences for
closed loops. It is likely that the $e_g$ electron mobility in Mn-oxides
will be reduced due to this quantum mechanical effect. Actually, the
double exchange model is recovered from Eqs.(9,10) only if
the absolute value of $t_{\bf lm}$ is considered since the following
identity holds:
$ |t_{\bf lm}| = 
\sqrt{ (1+\cos(\theta_{\bf l})\cos(\theta_{\bf m}) + \sin(\theta_{\bf l})
\sin(\theta_{\bf m})\cos(\phi_{\bf m}-\phi_{\bf l}) )/2} \\
=\cos(\theta_{\bf lm}/2)$, where $\theta_{\bf lm}$ is the angle between 
the core
spins at ${\bf l}$ and ${\bf m}$.\cite{zaanen}

To illustrate the importance of keeping the complex nature of the
hopping amplitude
we have studied numerically
Hamiltonian Eq.(9) with both $t_{\bf lm}$ and
$|t_{\bf lm}|$ for the particular cases of 1 and 2 electrons moving on a
$2 \times 2$ plaquette (4 sites on a square). Here in principle we need
8 angles to characterize a core spin configuration, but 3 of them
correspond to a rigid rotation of the whole system. Then, to study this
problem only 5 angles are needed. The four(six)
eigenvalues of Eq.(9) corresponding to one(two) particles can be found 
exactly for an arbitrary set of angles, and we average over about 2 million
randomly selected core spin configurations. Using this procedure we
calculated the
spectral density i.e. the normalized probability distribution for the
eigenvalues. The results are shown in Fig.1a-b. For the case where the
proper hopping Eq.(10) is used, a large accumulation of weight appears
at low energy (Fig.1a) where the spectral density grows sublinearly.
On the other hand, when only the absolute value of the
hopping is used in Fig.1b, the spectral density grows only linearly.
To the extent that the results for the square plaquette are qualitatively
representative of the bulk limit and other dimensions, 
the low energy accumulation of states could destabilize the ferromagnetic
ground state at relatively low 
temperatures in model Eq.(9) with a complex hopping amplitude.\cite{waves}
This is reasonable since the phase of the hopping
amplitude produces an effective random
magnetic field, which  could induce low energy modes 
and even localization effects in the bulk\cite{previous} 
in the phase competing with ferromagnetism
at $T>T_c$. Recent work in one dimensional NiO models has indeed shown
that the bandwidth of one hole at half-filling
is much reduced moving from a
Cu-background (spin 1/2) to a Ni-background (spin 1).\cite{nio}
In other words, while both model
Eqs.(9,10) for classical core spins and model Eq.(8) for
the S=2 systems favor the presence of a ferromagnetic state in the
ground state (zero temperature), 
they will likely differ with the double exchange model 
in the value of $T_c$.
Texture states of low energy may have such a large
degeneracy that the ferromagnetic state could be rapidly
destabilized with temperature.\cite{lynn} Thus, contrary to Ref.\cite{millis}
we believe that the Kondo-like model Eq.(1) may still describe the
physics of the Mn-oxides. This issue deserves further work.

Summarizing, we have derived a one band electronic model for the Mn-oxides
valid in the limit of a large Hund's coupling. The model is the analog
of the t-J model used for Cu-oxides, and of a
recently introduced model for NiO chains.\cite{nio} 
A generalization for arbitrary
spin S is given. At large S, the electrons move effectively with a
complex hopping amplitude.
Our Hamiltonian differs in a fundamental way from the double
exchange model in that the electrons can now acquire nontrivial
phases on closed loops, favoring interference,
localization and a small ferromagnetic $T_c$. We believe
that future theoretical work for Mn-oxides should use the proper
complex hopping amplitude Eq.(9) rather than the double exchange model, or,
even better, use the finite $S$ model Eqs.(6,8) proposed in this 
paper.

\medskip 
We thank A. Moreo, N. Bonesteel, E. Miranda and P. Wurth for discussions.
E. D. is supported by grant NSF-DMR-9520776.
E. M.-H. was supported in part through   
"Sonderforschungsbereich 341, Deutsche Forschungsgemeinschaft". 
We thank the National High Magnetic Field Lab for additional support.

\medskip

\vfil\eject

%
%

{\bf Figure Captions}

\begin{enumerate}


\item Spectral density $\rho(E)$ of non-vanishing energies
vs energy $E$ (in units of $t$)
in the large S limit of model Eq.(9) working on a
$2\times 2$ plaquette with two electrons. The result is an average over
2 million core spin angle configurations: (a) using the proper 
hopping amplitude Eq.(10); (b) using $|t_{\bf lm}|$ (double exchange).

\end{enumerate}


\begin{references}
\bibitem{sign} This sign is not contained in the Hamiltonian matrix (A6) in the
Appendix of Ref.\cite{anderson}.
\bibitem{spin} The polynomial for spin 3/2 is
 $Q_{S=3/2}(y) = - {{1}\over{2}} + {{3}\over{4}}y +
{{3}\over{4}} y^2$. Expressions for ${\rm S>2}$ can be easily obtained 
from the recursive formula Eq.(7).
\bibitem{waves} Note, however, that the spin waves are not effected by the 
Berry phase.
\end{references}
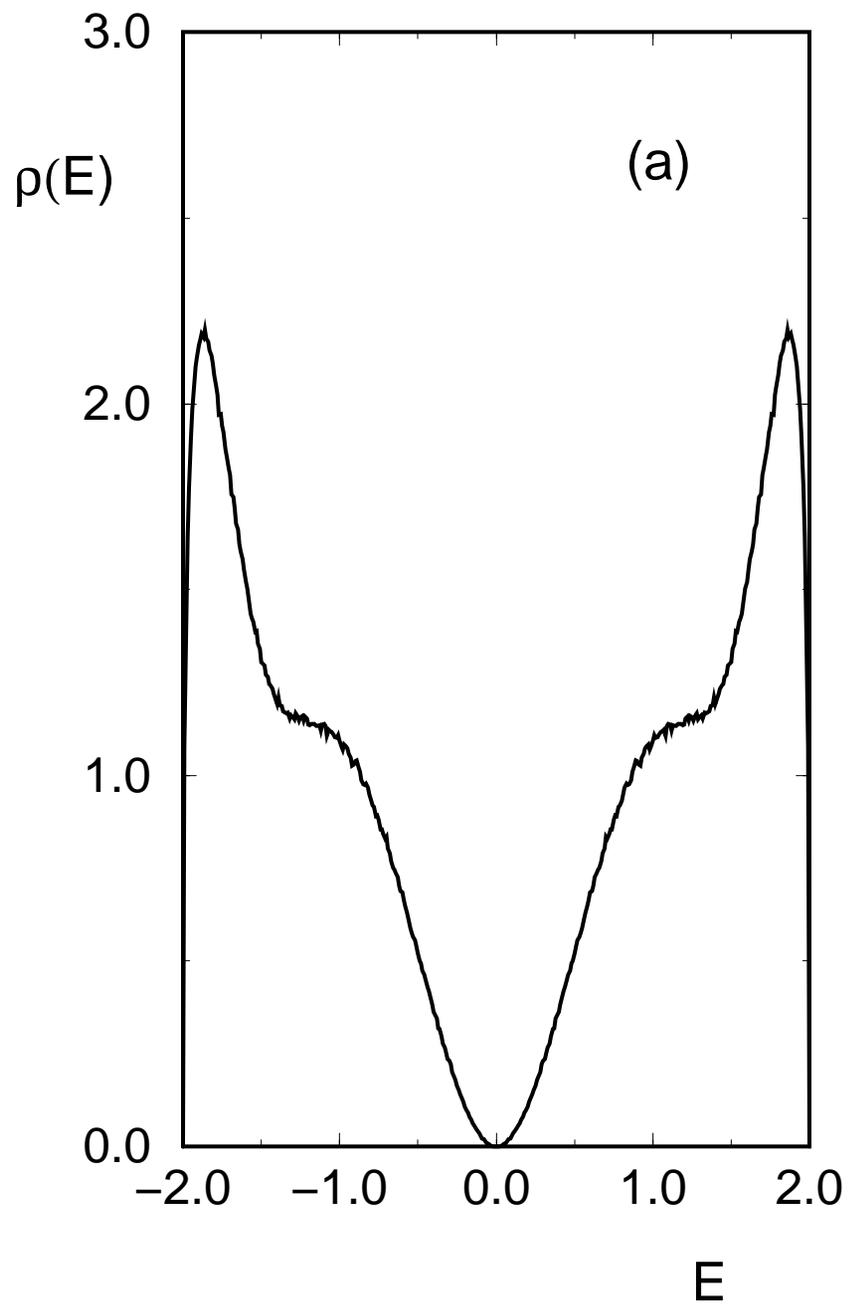
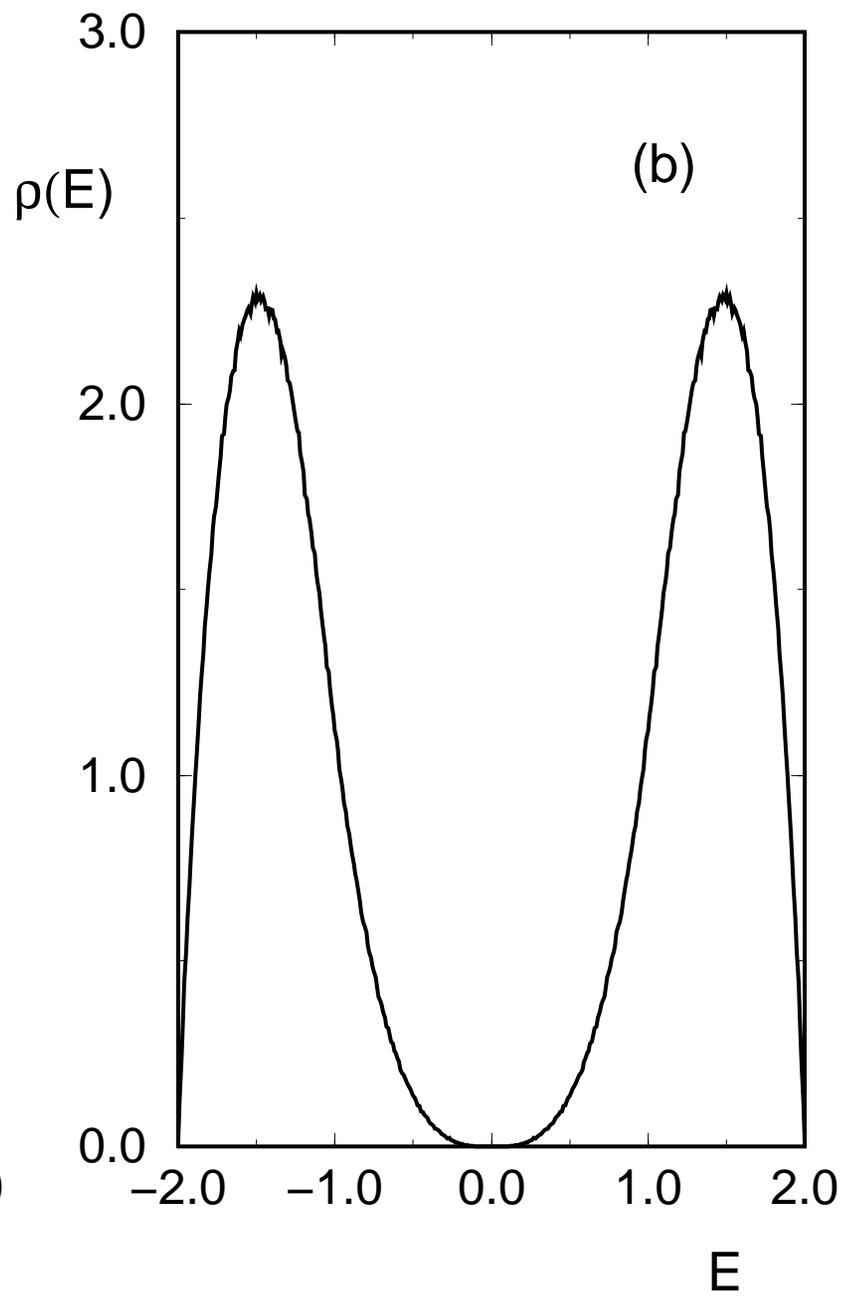
\end{document}